\documentclass[twocolumn]{aastex631}

\usepackage{CJK}
\usepackage{amsbsy}
\newcommand{\vect}[1]{\boldsymbol{#1}}

\accepted{May 25, 2021}

\shorttitle{Influence of the Hyperparameters on the Reconstruction of $H_0$ with GP}
\shortauthors{Sun et al.}
\graphicspath{{./}{figures/}}

\begin{document}
\begin{CJK*}{UTF8}{gbsn}
\title{Influence of the Bounds of the Hyperparameters on the Reconstruction of Hubble Constant with Gaussian Process}

\author[0000-0003-3995-4859]{Wen Sun (孙文)}
\affiliation{Department of Astronomy, Beijing Normal University, Beijing 100875, China}

\author{Kang Jiao (焦康)}
\affiliation{Department of Astronomy, Beijing Normal University, Beijing 100875, China}

\author{Tong-Jie Zhang (张同杰)}
\affiliation{Department of Astronomy, Beijing Normal University, Beijing 100875, China}
\affiliation{Institute for Astronomical Science, Dezhou University, Dezhou 253023, China}

\correspondingauthor{Tong-Jie Zhang}
\email{tjzhang@bnu.edu.cn}



\begin{abstract}

The cosmological model-independent method Gaussian process (GP) has been widely used in the reconstruction of Hubble constant $H_0$, and the hyperparameters inside GP influence the reconstructed result derived from GP. Different hyperparameters inside GP are used in the constraint of $H_0$ derived from GP with observational Hubble parameter $H(z)$ data (OHD), and the influence of the hyperparameters inside GP on the reconstruction of $H_0$ with GP is discussed. The discussion about the hyperparameters inside GP and the forecasts for future data show that the consideration of the lower and upper bounds on the GP's hyperparameters are necessary in order to get an extrapolated result of $H_0$ from GP reliably and robustly.

\end{abstract}

\keywords{Hubble constant(758), Observational cosmology(1146), Astronomy data analysis(1858), Computational methods (1965)}


\section{Introduction} \label{sec:intro}

The Hubble constant ($H_0$) characterizes the expansion rate of the universe at the current time, which is one of the most important parameters in cosmology. With the development of methods and technology, the measurement of the $H_0$ has become more and more accurate, but the $H_0$ tension crisis emerges. The locally and directly measured value of $H_0$ from \citet{2021ApJ...908L...6R} is $H_{0}=73.2\pm1.3$ km s$^{-1}$ Mpc$^{-1}$ (R21), while Planck Collaboration's work which is based on the Cosmic Microwave Background (CMB) observations and $\Lambda CDM$ model leads to a derived value of $H_{0}=67.4\pm0.5$ km s$^{-1}$ Mpc$^{-1}$ (P18) \citep{2020A&A...641A...6P}. The estimated error bars of these two results mentioned above do not overlap, and there is a 4.2$\sigma$ discrepancy between the two independent results. The $\Lambda CDM$ model which is a concordance model in cosmology is challenged by the tension between the two measurements of $H_0$. One of the potential explanations for this crisis is that the systematic errors that we do not know yet are influencing the two results \citep{2017NatAs...1E.169F, PhysRevD.103.063529}, but the real reasons for the discrepancy between the two independent results are still unknown. Therefore, the $H_0$ tension crisis becomes one of the most topical subjects in recent years, and the community attempts to use other independent methods to get a new derived value of $H_0$ in order to verify the existence of $H_0$ tension crisis. For example, \citet{2020arXiv200800487L} explore the cosmic expansion history with PAge approximation, which covers a broad class of cosmological models. They run Monte Carlo Markov Chain (MCMC) calculations with data from different sources and obtain $H_{0}=70.7\pm1.9$ km s$^{-1}$ Mpc$^{-1}$, which is consistent with both the local measurement and the CMB-inferred value. Their results show that it is hard to find a viable model which can resolve the $H_0$ tension with other ongoing debate in cosmology simultaneously.

As $H_0$ is the Hubble parameter at redshift $z=0$, it is easy to connect the measurement of $H_0$ with observational $H(z)$ data (OHD). The OHDs are the measurements of $H(z)$ at different redshifts which we can get by the cosmic chronometers (CC) or the radial baryon acoustic oscillations (BAO) data. One of the advantages of OHD is that the OHD from CC is cosmological model-independent, which means that we can get a model-independent measurement of $H_0$ with OHD from CC. \citet{2014MNRAS.441L..11B} use the non-parametric method Gaussian process (GP) to extrapolate OHD to $z=0$ in order to provide a model-independent measurement of $H_0$, and they obtain $H_{0}=64.9\pm4.2$ km s$^{-1}$ Mpc$^{-1}$ (1$\sigma$), which is in agreement with the CMB-inferred value. The OHD from CC should be paid more attention to as it is the only measurement based on the physics of the late universe that leads to a derived value of $H_0$ which is consistent with the CMB-inferred value. In the previous works about using GP with OHD to get model-independent measurements of $H_0$, they have already focused on the impact of different covariance functions inside GP and different OHD datasets on their measurements of $H_0$ \citep{2014MNRAS.441L..11B, 2018JCAP...10..015H, 2017SCPMA..60k0411W}, but they have not paid much attention to the influence of the lower and upper bounds of the hyperparameters inside GP on the extrapolated result of $H_0$ derived from GP.

In this paper, based on many previous works about using GP with OHD to get model-independent measurements of $H_0$, we move on to talk about the influence of hyperparameters inside the GP on the final measurement of $H_0$. We focus on whether measurements of $H_0$ derived from GP with different lower and upper bounds on GP's hyperparameters are still in agreement with the CMB-inferred value or not. With the lower and upper bounds on GP's hyperparameters which we set in Sec. \ref{sec:constraints}, we got measurements of $H_0$ which are not in agreement with the CMB-inferred value. We discuss our results in detail in Sec. \ref{sec:constraints}. The results show that the adjustments of the lower and upper bounds on the GP's hyperparameters are necessary when the GP is used to get an extrapolated result of $H_0$ precisely and reliably.

The paper is organized as follows: in the next section, we describe GP which is the method that we adopt to constrain $H_0$. In Sec. \ref{sec:constraints}, the measurements of $H_0$ with different lower and upper bounds on GP's hyperparameters and our discussion are displayed, followed by forecasts of constraints with simulated data set of $H(z)$ in Sec. \ref{sec:forecasts}. We finish the paper in Sec. \ref{sec:conclusions} with the conclusions.

\section{Gaussian Processes (GP)} \label{sec:method}

The GP is a collection of random variables, any finite number of which have a joint Gaussian distribution \citep{2006gpml.book.....R}. Gaussian process can be used to reconstruct $\vect{f^*}= \left\{f(z^{*}_i) \right\}$ at $\vect{Z^*}=\left\{ z^{*}_i \right\}$ from $n$ data points $\vect{y}= \left\{y_i \right\}$ at $\vect{Z}=\left\{ z_i \right\}$ without assuming parameterized forms of the function, which means that we can reconstruct the Hubble parameter $H(z)$ as a function of the redshift $z$ in a model-independent way with GP. The GP's reconstructed function $f(z)$ has the value which is a Gaussian random variable at a reconstructed point $z$, and a GP is completely confirmed by its mean function and covariance function. The function values at different reconstructed points are not independent of the others, and GP use the covariance function $cov(f(z), f(\tilde{z})) = k(z,\tilde{z})$ to describe the connection between the function values at different reconstructed points.

There are lots of covariance functions that we can choose depending on the type of problem that we are facing \citep{2006gpml.book.....R, 2013arXiv1311.6678S}. As we expect that covariance function should only depend on the distance between different points for simplicity, and the Hubble parameter should change smoothly with redshift, we decide to use the squared exponential covariance function, which has been widely used in this problem \citep{2014MNRAS.441L..11B, 2012JCAP...06..036S}. The squared exponential covariance function is given by
\begin{equation}
k\left(z,\tilde{z}\right)=\sigma_f^2exp\left(-\frac{{(z-\tilde{z})}^2}{2l^2}\right) .
\end{equation}

This covariance function depends on two ‘hyperparameters’ $\sigma_f$ and $l$, and the ‘hyperparameters’ means that $\sigma_f$ and $l$ are parameters of a non-parametric model. The length-scale $l$ determines the length in $z$-direction which corresponding to a significant change of $f(z)$, we can also understand the length-scale $l$ as how far do we need to move along the $z$-axis for the $f(z)$ values to be uncorrelated. While $\sigma_f$ determines the typical change of $f(z)$, which can also be seen as the amplitude or magnitude of the function.

By following the steps in \citet{2012JCAP...06..036S}, the mean $\overline{\vect{f^*}}$ and the covariance $cov(\vect{f^*})$ of the vector $\vect{f^*}= \left\{f(z^{*}_i) \right\}$ of the reconstructed function values at $\vect{Z^*}=\left\{ z^{*}_i \right\}$ is given by
\begin{equation}
\overline{\vect{f^*}}=\vect{\mu^*}+K(\vect{Z^*}, \vect{Z})[K(\vect{Z}, \vect{Z})+C]^{-1}(\vect{y}-\vect{\mu}) \label{con:mu}
\end{equation}
and
\begin{eqnarray}
cov(\vect{f^*})=&K(\vect{Z^*}, \vect{Z^*})&-\nonumber\\&K(\vect{Z^*}, \vect{Z})&[K(\vect{Z}, \vect{Z})+C]^{-1}K(\vect{Z}, \vect{Z^*}), \label{con:var}
\end{eqnarray}
respectively. Here $\vect{\mu}(\vect{z})$ is the a prior mean function, which is commonly set to zero, $\vect{\mu}(\vect{z})=0$ \citep{2018ApJ...856....3Y, 2012JCAP...06..036S, 2021EPJC...81..127B, 2020arXiv201110559R}. The $\vect{\mu}(\vect{z})$ in our work is set to zero as well. The covariance matrix $K(\vect{Z}, \vect{Z})$ is given by $[K(\vect{Z}, \vect{Z})]_{ij}=k(z_i, z_j)$, and $C$ is given by $C=diag(\sigma_{i}^2)$ for uncorrelated data corresponding to the Gaussian noise, where $\sigma_{i}^2$ is the variance of the Gaussian noise. The hyperparameters $\sigma_f$ and $l$ can be optimized and determined by maximizing the log marginal likelihood, which is given by
\begin{eqnarray}
\ln{\mathcal{L}}=&-\frac{1}{2}(\vect{y}-\vect{\mu})^T[K(\vect{Z}, \vect{Z})+C]^{-1}(\vect{y}-\vect{\mu})-\nonumber \\&\frac{1}{2}\ln{|K(\vect{Z}, \vect{Z})+C|}-\frac{n}{2}\ln{2\pi}. \label{con:maxlml}
\end{eqnarray}
The hyperparameters of GP are being fixed after being optimized through Eq.(\ref{con:maxlml}), then the expected value function and the variance of $\vect{f^*}$ can be calculated from Eq.(\ref{con:mu}) and Eq.(\ref{con:var}) respectively \citep{2012JCAP...06..036S,2021arXiv210108565C}. For the reconstruction of $H(z)$, as $H_0$ corresponds to $H(z)$ at $z=0$, the reconstructed result for $H_0$ can be obtained from the reconstructed $H(z)$ sequence. For further work which needs the reconstructed result from GP, the proper way of doing the analysis would be to treat the hyperparameters of GP as the same as the cosmological parameters, which means that vary them together and sample the joint posterior \citep{2021arXiv210402485D}.

As the hyperparameters characterize the ‘wiggles’ of the function, the lower and upper bounds on the two hyperparameters influence the reconstructed results derived from GP. As we only have rather vague information about the hyperparameters in GP, the influence of lower and upper bounds on the two hyperparameters of GP on the final reconstructed results should be considered. We determine the maximum likelihood value of the two hyperparameters $\sigma_f$ and $l$ with different lower and upper bounds on the two hyperparameters in order to obtain the reconstructed results and discuss the influence of lower and upper bounds on the two hyperparameters on the final measurements of $H_0$. We use Scikit-learn\footnote{\url{https://scikit-learn.org/stable/index.html}} which is a machine learning package in Python to achieve the GP in our work. In Scikit-learn package constant-value and length-scale are corresponding to $\sigma_f^2$ and $l$ respectively.

\section{Constraints on $H_0$} \label{sec:constraints}

We have collected 43 $H(z)$ measurements from the CC and the BAO respectively. 31 of them are from the CC \citep{2003ApJ...593..622J, 2005PhRvD..71l3001S, 2010JCAP...02..008S, 2012JCAP...07..053M, 2014RAA....14.1221Z, 2015MNRAS.450L..16M, 2016JCAP...05..014M, 2017MNRAS.467.3239R} while 12 of them are from the BAO \citep{2009MNRAS.399.1663G, 2012MNRAS.425..405B, 2012MNRAS.426..226C, 2013MNRAS.431.2834X, 2013MNRAS.429.1514S, 2013AA...552A..96B, 2014MNRAS.439...83A, 2014JCAP...05..027F, 2015AA...574A..59D}. The $H(z)$ measurements from CC are measured directly with spectroscopic dating techniques of passively-evolving galaxies and the assumption of a specific stellar population synthesis techniques (SPS) model. They are independent of the Cepheid distance scale and cosmological model-independent \citep{2018JCAP...04..051G, 2002ApJ...573...37J}. While the other $H(z)$ measurements from BAO are measured by using the anisotropy of the BAO signal, and they are dependent on cosmological model \citep{2018JCAP...04..051G, 2013MNRAS.431.2834X}. As the model-independent measurement of $H_0$ is more significant in the discussion about $H_0$ tension, we choose to use the 31 $H(z)$ measurements from CC to perform GP without using the other 12 $H(z)$ measurements from BAO in order to avoid the influence of their cosmological model dependence on the extrapolated results of $H_0$.

\begin{figure}[ht!]
\plotone{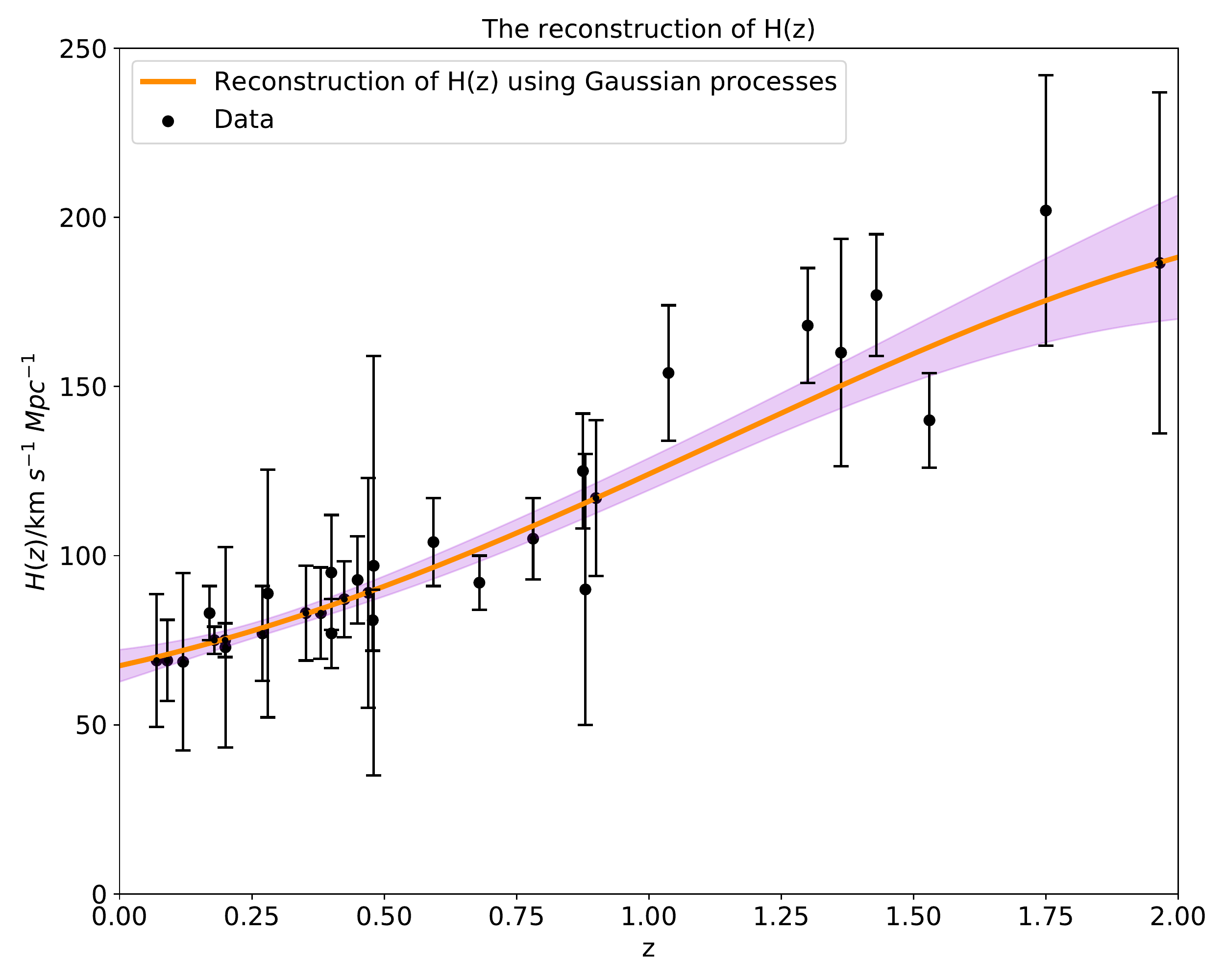}
\caption{The model-independent reconstructed result of $H(z)$ using Gaussian processes. The solid orange line and the pink shaded contour refer to the reconstructed result and the 1$\sigma$ confidence level respectively. The black points with error bars represent the 31 $H(z)$ measurements that we use for the reconstruction. Our reconstruction of $H(z)$ and extrapolated value of $H_0$ is completely independent of cosmological models. The measured value of $H_0$ that we get by GP is consistent with the P18. \label{fig:thebest_new_noprior}}
\end{figure}

According to the introduction about hyperparameters inside GP which have been discussed in the Sec. \ref{sec:method}, the rational ranges of the hyperparameters can be determined. As $H(z)$ is reconstructed in the range $0\le z\le2.0$, the reasonable estimate of the distance that we need to move along the $z$-axis for the $H(z)$ values to be uncorrelated is about 2. Based on the values of the 31 $H(z)$ measurements which are used for the reconstruction, the reasonable estimate of the amplitude of the reconstructed function is about 100. With the discussion above and the definitions of the hyperparameters in Sec. \ref{sec:method}, the rational ranges of the constant-value $\sigma_f^2$ and length-scale $l$ which are hyperparameters of squared exponential covariance function for our work about Hubble parameter are from 500 to 29500 and from 0.1 to 4.9 respectively. Without more restrictions imposing on the two hyperparameters, we finish the reconstruction and extrapolation of Hubble parameter with GP in the above region in hyperparameter space and obtain the reconstructed result for $H_0$ as $H_{0}=67.4\pm4.8$ km s$^{-1}$ Mpc$^{-1}$ (1$\sigma$). The reconstruction for Hubble parameter with GP is shown in Figure \ref{fig:thebest_new_noprior}. The solid orange line and the pink shaded contour refer to the reconstructed result and the 1$\sigma$ confidence level respectively. The black points with error bars represent the 31 $H(z)$ measurements which are used for the reconstruction. Our reconstruction of $H(z)$ and extrapolated value of $H_0$ is completely independent of cosmological models. The measured value of $H_0$ that we get by GP is highly consistent with the P18, while there is a 1.33$\sigma$ discrepancy between the result derived from GP and the R21. As our measured value of $H_0$ is model-independent while \textit{Planck}'s result is derived within $\Lambda CDM$ model, the comparison between our result and the P18 show that the measured value of $H_0$ derived from GP supports the $\Lambda CDM$ model. Our extrapolated result of $H_0$ is consistent with former results which are derived from OHD by GP as well \citep{2014MNRAS.441L..11B, 2018JCAP...04..051G}.

\begin{figure}[ht!]
\plotone{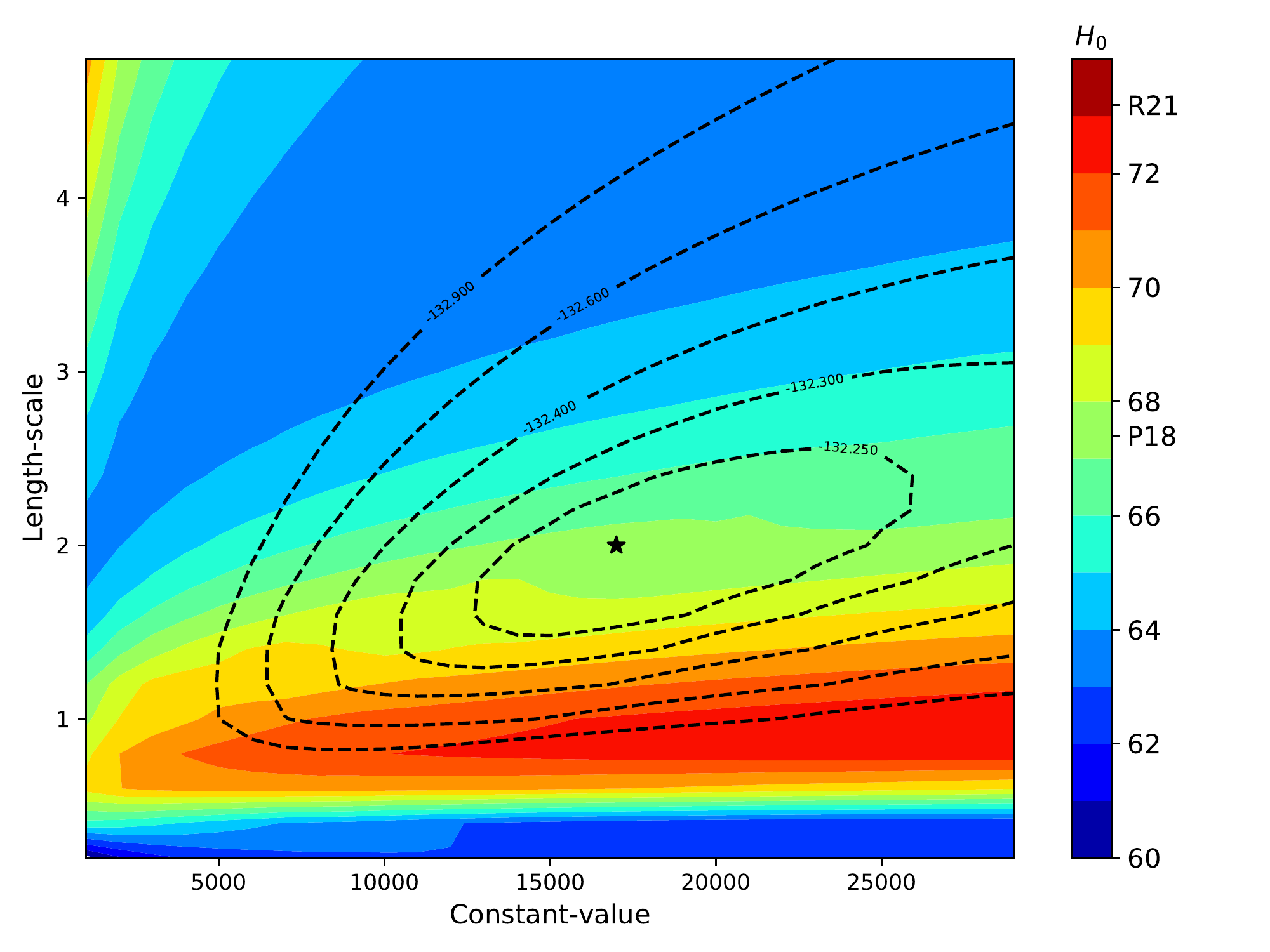}
\caption{The extrapolated results of $H_0$ using Gaussian processes on different pixels in hyperparameter space. The steps in the length-scale-axis direction and constant-value-axis direction of each rectangle in hyperparameter space are 0.2 and 1000, respectively. The contour plot with the color bar represents the extrapolated results of $H_0$ on each pixel in hyperparameter space, and the dashed lines represent the corresponding contour plot which shows the LML as a function of the two hyperparameters length-scale $l$ and constant-value $\sigma_f^2$. There is only one local maximum of LML in this range of hyperparameter space, indicated with the black star. \label{fig:result_heatwithLML}}
\end{figure}

In order to figure out the influence of lower and upper bounds on the two hyperparameters on the final measurements of $H_0$, we split the joint range of the two hyperparameters into grids in hyperparameter space and perform GP on each pixel in hyperparameter space. The steps in the length-scale-axis direction and constant-value-axis direction of each pixel in hyperparameter space are 0.2 and 1000, respectively. The results are shown in Figure \ref{fig:result_heatwithLML}. The contour plot with the color bar represents the extrapolated results of $H_0$ on each pixel in hyperparameter space, and the corresponding contour plot which shows the LML as a function of the two hyperparameters length-scale $l$ and constant-value $\sigma_f^2$ is shown in Figure \ref{fig:result_heatwithLML} as well in order to show the variation of the LML in hyperparameter space clear, indicated with the dashed lines. The point with a black star represents the local maximum of LML. The extrapolated value of $H_0$ has a significant correlation with the value of length-scale $l$. As shown in Figure \ref{fig:result_heatwithLML}, in hyperparameter space, most extrapolated results of $H_0$ on pixels are consistent with the P18 rather than the R21, especially in regions where have large LMLs. There are only two regions in this area that prefer the R21 rather than the P18. One is a small region in the upper left corner, while the other is a narrow space that lies in the constant-value-axis direction, corresponding to the region in hyperparameter space where length-scale $l\approx1$. As the first one corresponds to a small region with small LMLs, even if our lower and upper bounds on the two hyperparameters are close to the lower right corner, GP does not prefer their higher values of $H_0$ as well. For the second region near to the local maximum of LML, if our lower and upper bounds on the two hyperparameters are close to this region described above and do not contain the local maximum of LML, we will maybe get an extrapolated result of $H_0$ which is consistent with the R21. As shown in the lower panel of Figure \ref{fig:result_heatwithLML}, there is only one local maximum of LML in this range of hyperparameter space, which is great for our work. But in fact, there may exist more than one local maximum of LML in the range of hyperparameter space that we consider. As the gradient-based optimization may converge to any local maxima of LML in the range of hyperparameter space that we consider depending on the initial value for the hyperparameters, we should focus on the illustration of the LML landscape in the range of hyperparameter space that we consider, and adjust the lower and upper bounds on the hyperparameters in order to get the good local optimum. We think the consideration of the LML landscape and the adjustments of the lower and upper bounds on the hyperparameters together is more convenient and convincing than repeating the optimization with the same lower and upper bounds on the hyperparameters several times for different initial values of the hyperparameters.

We also compare the extrapolated results of $H_0$ using GP on different pixels in hyperparameter space with other methods' results. Firstly, if the reconstruction for Hubble parameter with GP on one pixel in hyperparameter space does not satisfy the physical condition $\frac{\mathrm{d}H}{\mathrm{d}z}>0$ which means that the background energy density decreases with time \citep{2020ApJ...892L..28H}, we will mask the corresponding pixel in Figure \ref{fig:result_heatwithLML}. Secondly, for the pixels in Figure \ref{fig:result_heatwithLML} which are not masked, if an extrapolated result of $H_0$ on one of them satisfy that both the center value of the measurement derived from GP and the center value of the P18 is in each other's 3$\sigma$ confidence level, we will regard as the result derived from GP is consistent with the P18. The comparison between the result derived from GP and the result derived from another method holds the same criterion as above. If an extrapolated result of $H_0$ on one pixel does not consistent with the result derived from another method, we will mask the corresponding pixel in Figure \ref{fig:result_heatwithLML} as well. We compare the result derived from GP with the P18, the R21, and the result derived within PAge approximation respectively. The results of comparisons are shown in Figure \ref{fig:result_heat_masked}. The region where the result derived from GP is consistent with the P18 in hyperparameter space is much larger than the region where the result derived from GP is in agreement with the R21 in hyperparameter space, and pixels in this region have large LML, which means that if we perform GP in one part of the hyperparameter space which we consider in our work, we will probably get an extrapolated result of $H_0$ which is consistent with the P18. There is a large region with large LML in hyperparameter space where the result derived from GP is consistent with the result derived within PAge approximation. As GP is model-independent, this comparison result means that it is worthwhile to focus on cosmological models like PAge approximation in the further discussion about the $H_0$ tension crisis.

\begin{figure*}
\gridline{\fig{result_heat_masked_cmb_1.pdf}{0.9\textwidth}{(a)}
          }
\gridline{\fig{result_heat_masked_local_1.pdf}{0.9\textwidth}{(b)}
          }
\gridline{\fig{result_heat_masked_PAge_1.pdf}{0.9\textwidth}{(c)}
          }
\caption{The results of comparison between the extrapolated result of $H_0$ derived from GP with different hyperparameters and the measurements of $H_0$ derived from other methods. (a), (b) and (c) show the results of comparison between the extrapolated result of $H_0$ derived from GP with CMB-inferred value, Riess's result, and the result derived from PAge approximation respectively. \label{fig:result_heat_masked}}
\end{figure*}

\section{Forecasts} \label{sec:forecasts}

In order to show the necessity of the consideration of the GP's hyperparameters more clearly, the next step is to focus on the LML landscape of GP in hyperparameter space with more data points. We use the method of \citet{Ma_2011} to generate the simulated data set of $H(z)$ with the recent measurements from CC \citet{2003ApJ...593..622J, 2005PhRvD..71l3001S, 2010JCAP...02..008S, 2012JCAP...07..053M, 2014RAA....14.1221Z, 2015MNRAS.450L..16M, 2016JCAP...05..014M, 2017MNRAS.467.3239R}. A simulated $H(z)$ value $H_{sim}(z)$ at any given redshift $z$ is consisted of the flducial value $H_{fid}(z)$ and a random deviation of $H_{sim}(z)$ from $H_{fid}(z)$ as $\Delta H$. The flducial value $H_{fid}(z)$ is given by
\begin{equation}
H_{fid}(z)=H_{0}\sqrt{\Omega_{m}(1+z)^3+\Omega_{\Lambda}},
\end{equation}
where $\Omega_{m}=0.27$, $\Omega_{\Lambda}=0.73$ and $H_0$ is the Hubble constant.

\begin{figure}[ht!]
\plotone{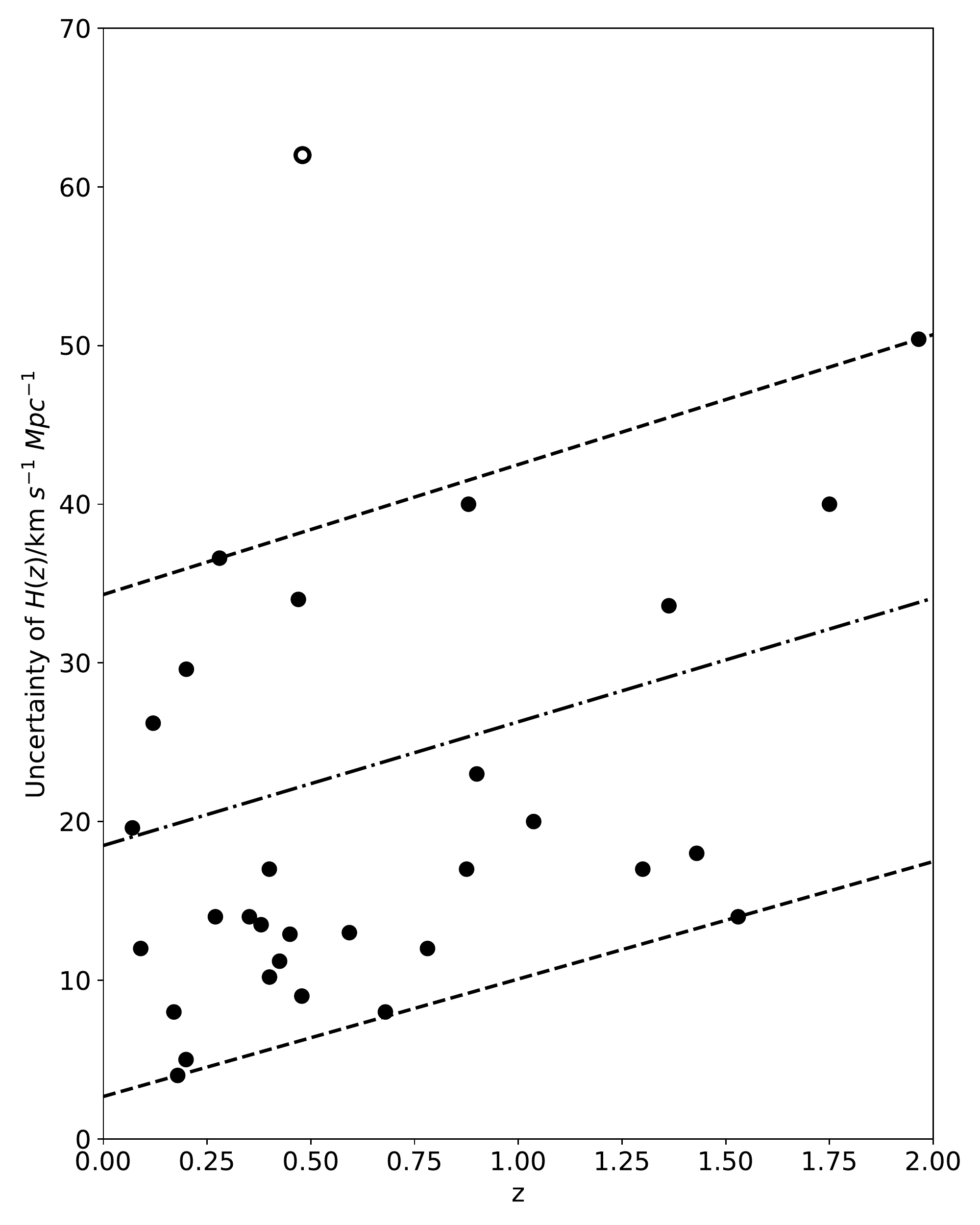}
\caption{The uncertainties of the current 31 $H(z)$ measurements. The circle and solid dots represent outlier and non-outliers, respectively. The bounds $\sigma_{+}$ and $\sigma_{-}$ are plotted as the two dashed lines. The dash-dotted line represents estimated mean uncertainty $\sigma_{0}$. \label{fig:H_sim}}
\end{figure}

The random deviation $\Delta H$ satisfies the Gaussian distribution $N(0, \tilde{\sigma}(z))$, in which $\tilde{\sigma}(z)$ is the uncertainty. This uncertainty is drawn by following the steps in \citet{Ma_2011} which are based on the assumption that the uncertainties of future data will follow the increasing trend with redshift $z$ of the current 31 $H(z)$ measurements. The uncertainties of the current 31 $H(z)$ measurements are shown in Figure \ref{fig:H_sim}. The uncertainties of the $H(z)$ measurements are restricted by two straight lines $\sigma_{+}=8.19z+34.31$ and $\sigma_{-}=7.40z+2.67$ except an outlier from the data set. We take the mean line of the strip $\sigma_{0}=7.80z+18.49$ as an estimate of the mean uncertainty of the simulated data. In this way, the uncertainty $\tilde{\sigma}(z)$ which is a random number can be drawn from the Gaussian distribution $N(\sigma_{0}(z), \epsilon(z))$, in which $\epsilon(z)=(\sigma_{+}-\sigma_{-})/4$ in order to make sure the probability of $\tilde{\sigma}(z)$ falling within the strip is 95.4\%. Finally, we can get the simulated $H(z)$ value $H_{sim}(z)=H_{fid}(z)+\Delta H$.

\begin{figure}[ht!]
\plotone{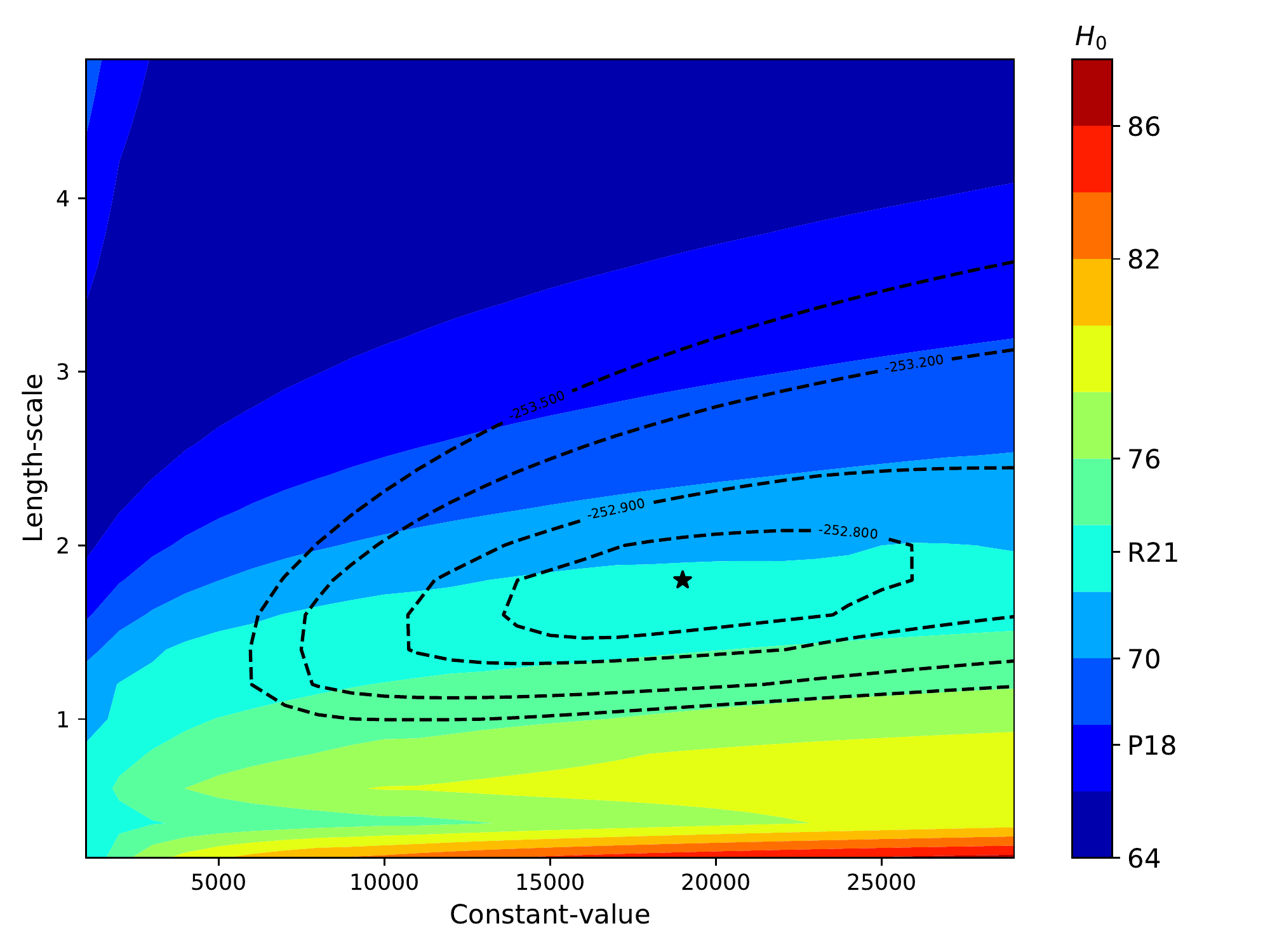}
\caption{The expected future extrapolated results of $H_0$ using Gaussian processes on different pixels in hyperparameter space with the simulated data points and the 31 $H(z)$ measurements from CC. The contour plot with the color bar represents the extrapolated results of $H_0$ on each pixel in hyperparameter space, and the dashed lines represent the corresponding contour plot which shows the LML as a function of the two hyperparameters length-scale $l$ and constant-value $\sigma_f^2$. The local maximum of LML is indicated with the black star. \label{fig:heatandLML_sim}}
\end{figure}

With the method above, we generate a simulated data set of 25 data points that are equally spaced in the range $0.1\le z\le2.0$. With the simulated data points and the 31 $H(z)$ measurements from CC, Figure \ref{fig:heatandLML_sim} presents the expected future extrapolated results of $H_0$ using GP and the corresponding LML landscape of GP in hyperparameter space. The contour plot with the color bar represents the extrapolated results of $H_0$ on each pixel in hyperparameter space, and the dashed lines represent the corresponding contour plot which shows the LML as a function of the two hyperparameters length-scale $l$ and constant-value $\sigma_f^2$. The local maximum of LML is indicated with the black star. For the future data set with the same quality as today and the 31 $H(z)$ measurements from CC, the LML landscape of GP changes. There exist more regions with high LMLs in which the extrapolated results of $H_0$ using GP are consistent with the R21 or the P18. In this situation, it is more important to focus on the ranges of the hyperparameters in order to determine $H_0$ precisely and reliably.

\section{Conclusions and Discussions} \label{sec:conclusions}

We make the reconstruction and extrapolation of the Hubble parameter with the GP. The reconstruction of $H(z)$ and extrapolated value of $H_0$ is completely independent of cosmological models, which means that our results can be used to discuss the $H_0$ tension crisis. We obtain the reconstructed result for $H_0$ as $H_{0}=67.4\pm4.8$ km s$^{-1}$ Mpc$^{-1}$ (1$\sigma$), which is remarkably consistent with the P18. This supports that either there are systematic errors that we do not know yet in Riess's measurement of $H_0$, or there is a difference between the local value and the global value. Our result is larger than result from \citet{2014MNRAS.441L..11B} with 19 $H(z)$ measurements and very close to the result from \citet{2017SCPMA..60k0411W} with 30 $H(z)$ measurements. This comparison shows that the recent new $H(z)$ measurements make the extrapolated result of $H_0$ from GP become larger and more consistent with the P18.

More importantly, the influence of lower and upper bounds of the hyperparameters of GP on the final measurements of $H_0$ is discussed in order to show that it is necessary to consider the lower and upper bounds on the hyperparameters before we get an extrapolated result of $H_0$ from the GP precisely and reliably. The rational ranges of the hyperparameters should be determined based on the estimate of the correlation between the function values at different reconstructed points, and the estimate of the amplitude or magnitude of the reconstructed function. In order to confirm whether there are more than one local maxima of LML in the range of hyperparameter space that we consider or not, we should check out the LML landscape of GP as well. We should determine which local maximum is the best one based on values of the hyperparameters and characteristics of the reconstructed functions corresponding to different local maxima in hyperparameter space. Then we can adjust the lower and upper bounds on the hyperparameters in order to make sure the gradient-based optimization converges to the best local maximum of LML.

In the future, with more data points and different kernels of GP, the LML landscape of GP will change, which means that there may exist more than one local maximum of LML in the range of hyperparameter space that we consider, and there may exist more regions with high LMLs in which the extrapolated results of $H_0$ using GP is consistent with both the CMB-inferred value and local measurement of $H_0$. The reconstructed results of $H_0$ using GP may become more sensitive to the lower and upper bounds of the GP's hyperparameters. Forecasts have shown that in this situation the discussion and consideration of the lower and upper bounds on the GP's hyperparameters become more important in order to get an extrapolated result of $H_0$ from GP precisely and reliably.

\begin{acknowledgments}
We are very grateful to the referee for the useful comments. We are very grateful to Zhiqi Huang for his kind help and thank Keyu Xing and Junshuo Zhang for useful discussions. This work is supported by National Key R\&D Program of China (2017YFA0402600), and National Science Foundation of China (Grants No. 11929301, 11573006).
\end{acknowledgments}


\bibliography{GP_hyperparameters}{}
\bibliographystyle{aasjournal}


\end{CJK*}
\end{document}